\documentclass[aps,pra,twocolumn,groupedaddress,amsmath,amssymb,showpacs]{revtex4}
\usepackage{graphicx}
\usepackage{amsmath}
\usepackage{dcolumn}
\usepackage{bm}
\newcommand{\tr}{\mathrm{Tr}}

\begin{document}
\title{Entangled mixed-state generation by twin-photon scattering}
\author{G. Puentes$^1$}
\author{A. Aiello$^1$}
\author{D. Voigt$^{1,2}$}
\author{J.P. Woerdman$^1$}
\affiliation{1. Huygens Laboratory, Leiden University, P.O. Box
9504, 2300 RA Leiden, The
Netherlands. \\
2. Cosine Research bv, Niels Bohrweg 11, 2333 CA Leiden, The
Netherlands.}
\begin{abstract}
We report  experimental  results on mixed-state generation by
multiple scattering of polarization-entangled photon pairs created
from parametric down-conversion. By using a large variety of
scattering optical systems we have experimentally obtained
entangled mixed states that lie upon and below the Werner curve in
the linear entropy-tangle plane. We have also introduced a simple
phenomenological model built on the analogy between classical
polarization optics and quantum maps. Theoretical predictions from
such model are in full agreement with our experimental findings.
\end{abstract}


\maketitle



%
%
%
%
%
%
%
%
%
%
%
%
\section{Introduction}
The study of spatial, temporal and polarization correlations of
light scattered by inhomogeneous and turbid media has a history of
more than a century \cite{rayleigh}. Due to the high complexity of
scattering media only single-scattering properties are known at a
microscopic level \cite{hulst}. Conversely, for
multiple-scattering processes the emphasis is mainly on
macroscopic theoretical descriptions of the correlation phenomena
\cite{rossum}. In most examples of the latter
\cite{albada1,chabanov,rikken,freund} the intensity correlations
of the interference pattern generated by multiple-scattered light
are explained in terms of \emph{classical} wave-coherence. On the
other hand, the recent availability of reliable single-photon
sources has triggered the interest in \emph{quantum} correlations
of multiple-scattered light \cite{lohdal}. Generally speaking,
quantum correlations of  scattered photons  depend on the quantum
state of the light illuminating the sample. In Ref. \cite{lohdal},
\emph{spatial} quantum correlations of scattered light were
analyzed for Fock, coherent and thermal input states.

In this paper we present the first experimental results on quantum
polarization correlations of scattered photon pairs. Specifically,
we study the entanglement content in relation to the polarization
purity of multiple-scattered twin-photons, initially generated in
a polarization-entangled state  by spontaneous parametric
down-conversion (SPDC). The initial entanglement of the input
photon pairs will in general be degraded by  multiple scattering.
This can be understood by noting that the scattering process
distributes the initial correlations of the twin-photons over the
many spatial modes excited along the propagation in the medium. In
the case of spatially inhomogeneous media the polarization degrees
of freedom are coupled to the spatial degrees of freedom
generating polarization dependent speckle patterns. If the spatial
correlations  of such patterns are averaged out by multi-mode
detection, the polarization state of the scattered photon(s) is
reduced to a mixture, and the resulting polarization-entanglement
of the photon pairs is degraded with respect to the initial one. A
related theoretical background was elaborated in
\cite{aiello1,vanvelsen}.

This paper is structured as follows: In section II we report our
experiments on light scattering with entangled photons. First, we
present our experimental set-up and briefly describe the many
different optical systems that we used as scatterers. Next, we
show our experimental results. The notions of generalized Werner
and sub-Werner states are introduced to illustrate these results.
%
%
In section III we introduce a simple phenomenological model for
photon scattering that fully reproduces our experimental findings.
Finally, in section IV we draw our conclusions.
\section{Experiments on light scattering with entangled photons}
\subsection{Experimental set-up}
\begin{figure}[!h]
\includegraphics[angle=0,width=8truecm]{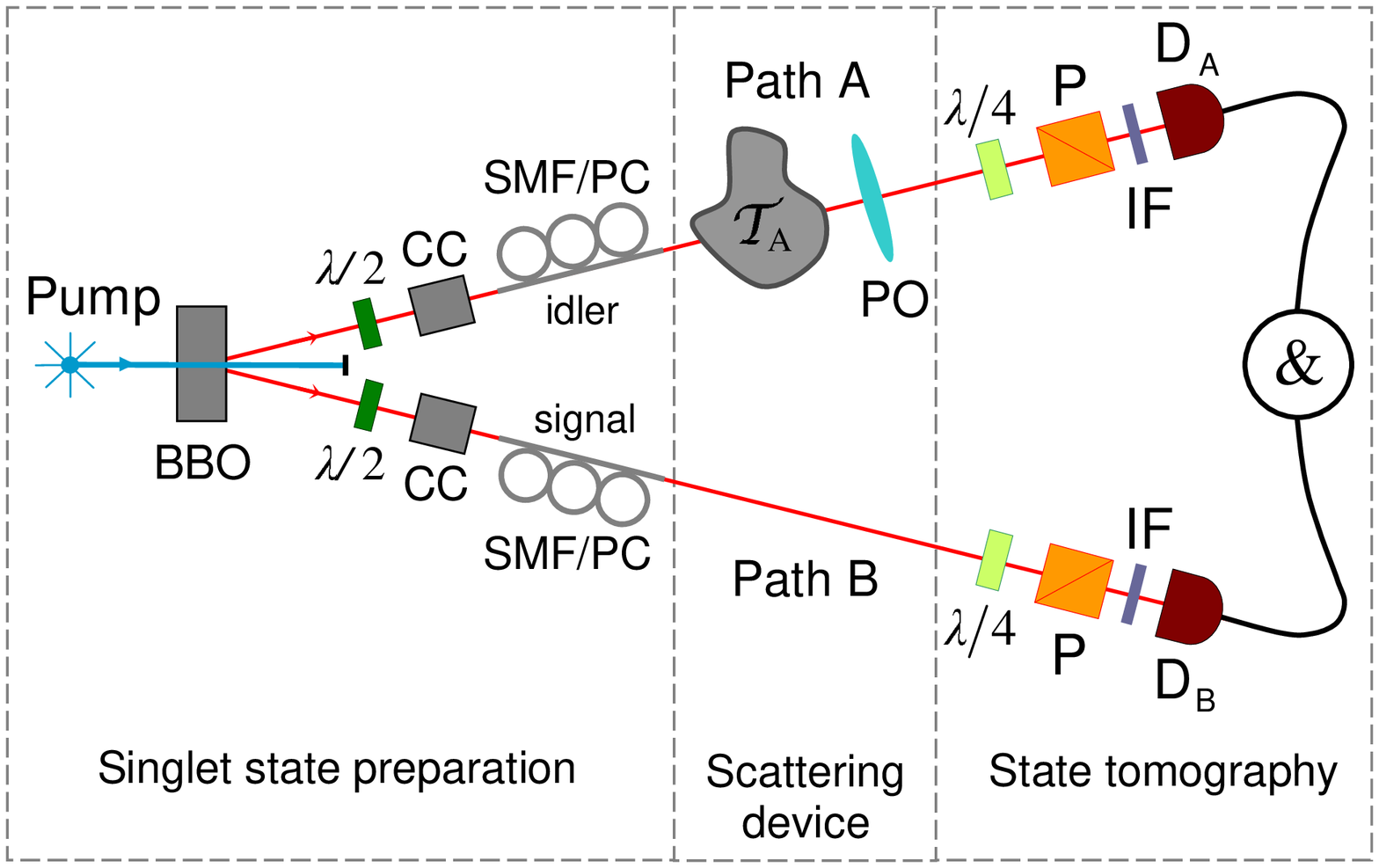}
\caption{\label{fig:1}(Color online) Experimental scheme: After
singlet preparation, the idler photon propagates through the
scattering system $\mathcal{T}_A$. The polarization state of the
scattered photon-pairs is then reconstructed via a quantum
tomographic procedure (see text for details).}
\end{figure}

%
%
Our experimental set-up is shown in Fig.~\ref{fig:1}.~A
Krypton-ion laser at 413.1~nm  pumps a 1~mm thick
$\beta-\mathrm{Ba} \mathrm{B}_2 \mathrm{O}_4$ (\textsf{BBO})
crystal, where polarization-entangled photon pairs at wavelength
826.2~nm are created by SPDC in degenerate type II phase-matching
configuration \cite{sergienko}. Single-mode fibers (\textsf{SMF})
are used as spatial filters to assure that each photon of the
initial SPDC pair travels in a single transverse mode. Spurious
birefringence along the fibers is compensated by suitably oriented
polarization controllers (\textsf{PC}). The total retardation
introduced by the fibers and walk-off effects at the \textsf{BBO}
crystal are compensated by compensating crystals (\textsf{CC}:
0.5~mm thick \textsf{BBO} crystals) and half-wave plates
($\lambda/2$), in both signal and idler paths. In this way the
initial two-photon state is prepared in the polarization singlet
state $|\psi_s \rangle=(|HV\rangle-|VH\rangle)/\sqrt{2}$, where
$H$ and $V$ are labels for horizontal and vertical polarizations
of the two photons, respectively.
The experimentally prepared initial singlet state
$\rho_s^\mathrm{exp}$ has a fidelity \cite{fidelity} with the
theoretical singlet state $\rho_s = | \psi_s \rangle \langle
\psi_s |$ of $F(\rho_s,\rho_s^\mathrm{exp}) \sim 98\%$.
In the second part of the experimental set-up the idler photon
passes though the scattering device $\mathcal{T}_A$ before being
collimated by a photographic objective (\textsf{PO}) with focal
distance $f = 5~\mathrm{cm}$.
The third and last part of the experimental set-up, consists of
two tomographic analyzers (one per photon), each  made of a
quarter-wave plate ($\lambda/4$) followed by a linear polarizer
(\textsf{P}). Such analyzers permit a full tomographic
reconstruction, via a maximum-likelihood  technique \cite{james},
of the two-photon state. Additionally, interference filters
(\textsf{IF}) put in front of each detector ($\Delta \lambda = 5$~
nm) provide for bandwidth selection. Detectors
$\mathrm{\textsf{D}}_\mathrm{\textsf{A}}$ and
$\mathrm{\textsf{D}}_\mathrm{\textsf{B}}$ are ``bucket''
detectors, that is they do not distinguish which spatial mode a
photon comes from, thus each photon is detected in a
\emph{mode-insensitive} way.
%
%
\subsection{Scattering devices}
All the scattering optical systems that we used were located in
the path of only \emph{one} of the photons of the entangled-pair
(the idler one), as shown in Fig.~\ref{fig:1}. For this reason, we
refer to such systems as  \emph{local} scatterers. Such scatterers
can be grouped in three general categories according to the
optical properties of the media they are made of \cite{puentes}:
\begin{description}
  \item[Type I]  Purely depolarizing media, or diffusers.  Such media do not affect directly
  the polarization state of the impinging light but change the
  spatial distribution of the impinging electromagnetic field.
  \item[Type II]  Birefringent  media, or retarders. These media  introduce
a polarization-dependent \emph{delay} between different components
of the electromagnetic field.
  \item[Type III] Dichroic media, or diattenuators. Such media  introduce
 polarization-dependent \emph{losses} for the different components of the
electromagnetic field.
\end{description}
Type I  scattering systems produce an isotropic spread in the
momentum of the impinging photons. Examples of such scattering
devices are: spherical-particle suspensions (such as milk or
polymer micro-spheres), polymer and glass multi-mode fibers and
surface diffusers.
 Type II  scattering systems are made of birefringent
 media, which introduce an optical axis that breaks polarization-isotropy.
Birefringence can be classified as  ``material birefringence''
when it is an intrinsic property of the bulk medium (for example a
birefringent wave-plate), and as ``topological birefringence''
when it is induced by a special geometry of the system that
generates polarization-anisotropy, an example of a system with
topological birefringence is an array of cylindrical particles.
Finally, type III scattering systems are made of dichroic media
that produce polarization-dependent photon absorbtion. Examples of
such devices are commonly used polarizers.
 A systematic characterization of all
the scattering devices that we used was given in Ref.
\cite{puentes}.
\subsection{ Experimental results  \\
in the tangle versus linear entropy plane }
The degree of entanglement and the degree of mixedness of the
scattered photon pairs can be quantified by the tangle ($T$),
namely, the concurrence squared \cite{wooters}, and the linear
entropy ($S_L$) \cite{vedral}. These quantities were calculated
from the $4 \times 4$ polarization two-photon density matrix
$\rho$, by using $T(\rho)$=(max$\{0,\sqrt{\lambda_{1}}-
\sqrt{\lambda_{2}}-\sqrt{\lambda_{3}}-\sqrt{\lambda_{4} }\})^2$,
where $\lambda_{1} \geq \lambda_{2} \geq\lambda_{3}
\geq\lambda_{4}\geq 0 $ are the  eigenvalues of $\rho(\sigma_{2}
\otimes \sigma_{2})\rho^{*}(\sigma_{2} \otimes \sigma_{2})$, where
$\sigma_2 = \left[ \begin{array}{cc}
  0 & -i \\
  i & 0
\end{array} \right]$, and
$S_{L}(\rho)=\frac{4}{3}[1-\mathrm{Tr}(\rho^2)]$.
%
%
\begin{figure}[ht]
\includegraphics[angle=0,width=7truecm]{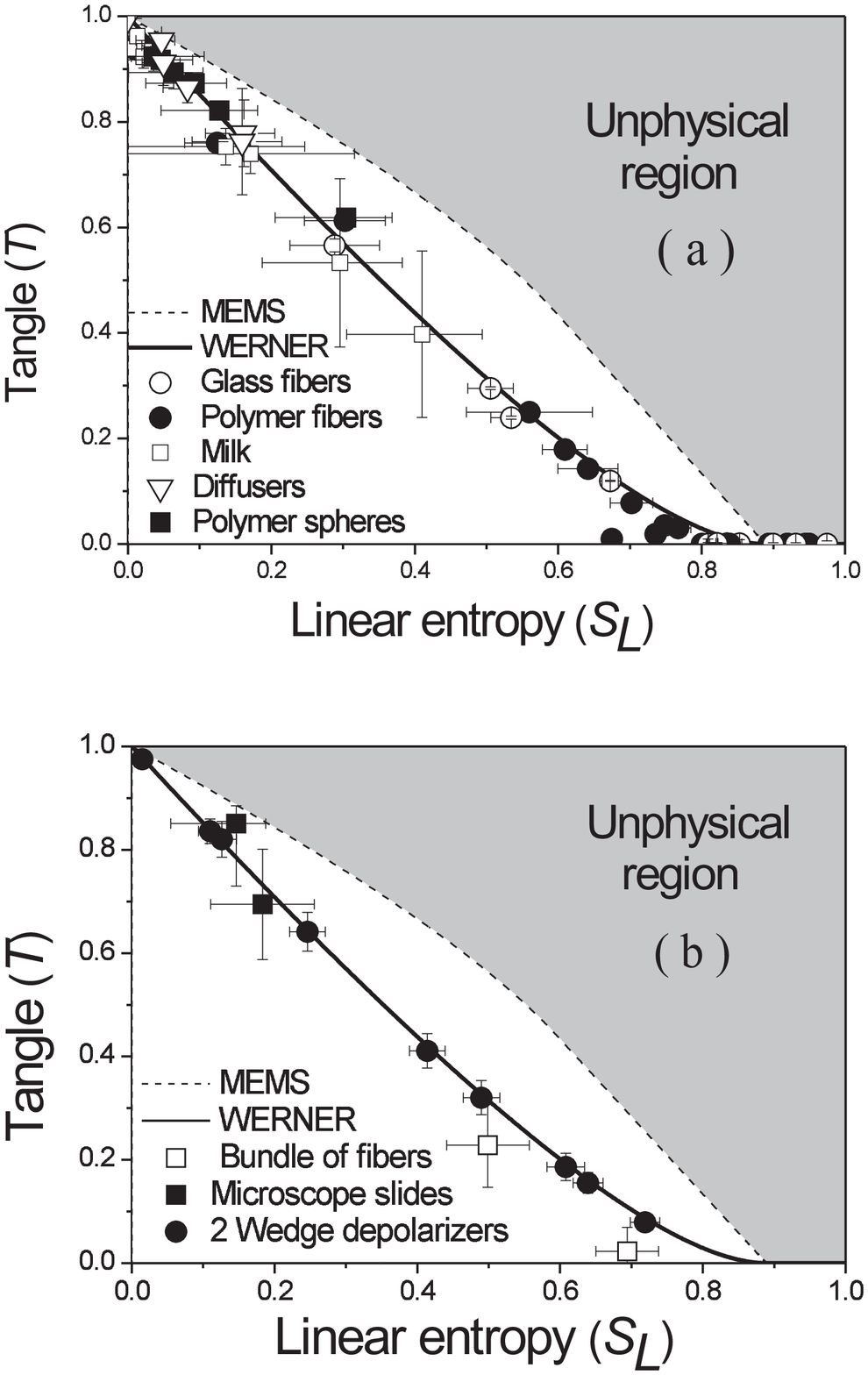}
\caption{\label{fig:2} Experimental data in the linear
entropy-tangle ($S_{L}$-$T$) plane. The grey area corresponds to
unphysical density matrices. Dashed upper curve: Maximally
entangled mixed states (MEMS), continuous lower curve: Werner
states. (a)~Polarization-isotropic scatterers (type I).
(b)~Birefringent scatterers (type II).}
\end{figure}
Figures~\ref{fig:2}~(a) and (b) show experimental data reported on
the linear entropy-tangle  plane. The position of each
experimental point in such plane has been calculated from  a
tomographically  reconstructed \cite{james} two-photon density
matrix $\rho^\mathrm{exp}$. The uniform grey area corresponds to
non-physical states \cite{mems}. The dashed curve that bounds the
physically admissible region from above is generated by the
so-called maximally entangled mixed states (MEMS)
\cite{mems2,mems3}. The lower continuous curve is produced by the
Werner states \cite{werner} of the form: $\rho_{W}=p \rho_s
+\frac{1-p}{4}I_4$, ($0\leq p \leq 1 $), where $I_4$ is the $4
\times 4$ identity matrix.~Figure~\ref{fig:2}~(a) shows
experimental data generated by isotropic scatterers (type I).
Specifically, our type I scatterers  consisted of the following
categories. (i) Suspensions of milk and micro-spheres in distilled
water, where the sample dilution was varied to obtain different
points; (ii) Multi-mode glass and polymer fibers, where the tuning
parameter exploited to obtain different points was the length of
the fiber (cut-back method); (iii) Surface diffusers, where the
full width scattering angle was used as tuning parameter. It
should be noted that suspensions of milk and micro-spheres are
\emph{dynamic} media, where Brownian motion of the micro-particles
induces temporal fluctuations within the detection integration
time \cite{puentes}.

In Fig.~2~(a), the experimental point at the top-left corner
(nearby $T=1$, $S_L=0$), is generated by the un-scattered initial
singlet state. The net effect of scattering systems with
increasing thickness is to shift the initial datum toward the
bottom-right corner ($T=0$, $S_L=1$), that corresponds to a fully
mixed state.

Figure~\ref{fig:2}~(b) displays experimental data generated by
birefringent scattering systems (type II). As an example of a
system with ``material birefringence'' we used a pair of wedge
depolarizers in cascade \cite{kliger}. Different experimental
points where obtained by varying the relative angle between the
optical axis of the two wedges \cite{puentes2}.
The systems with ``topological birefringence'' we considered
consisted of two different devices:
(i) The first one was a bundle of parallel optical fibers
\cite{vdmark}. Translational invariance along the fibers axes
restricts the direction of the wave-vectors of the scattered
photons in a plane orthogonal to the common axis of the fibers.
(ii) The second device was a stack of parallel microscope slides
(with uncontrolled air layers in between). This optical system is
depolarizing because it amplifies any initial spread in the
wave-vector of the impinging photon. This photon enters via a
single-mode-fiber (numerical aperture $\mathrm{NA} = 0.12$), from
one side of the stack and travels in a plane parallel to the
slides.

\begin{figure}[!h]
\includegraphics[angle=0,width=6.5truecm]{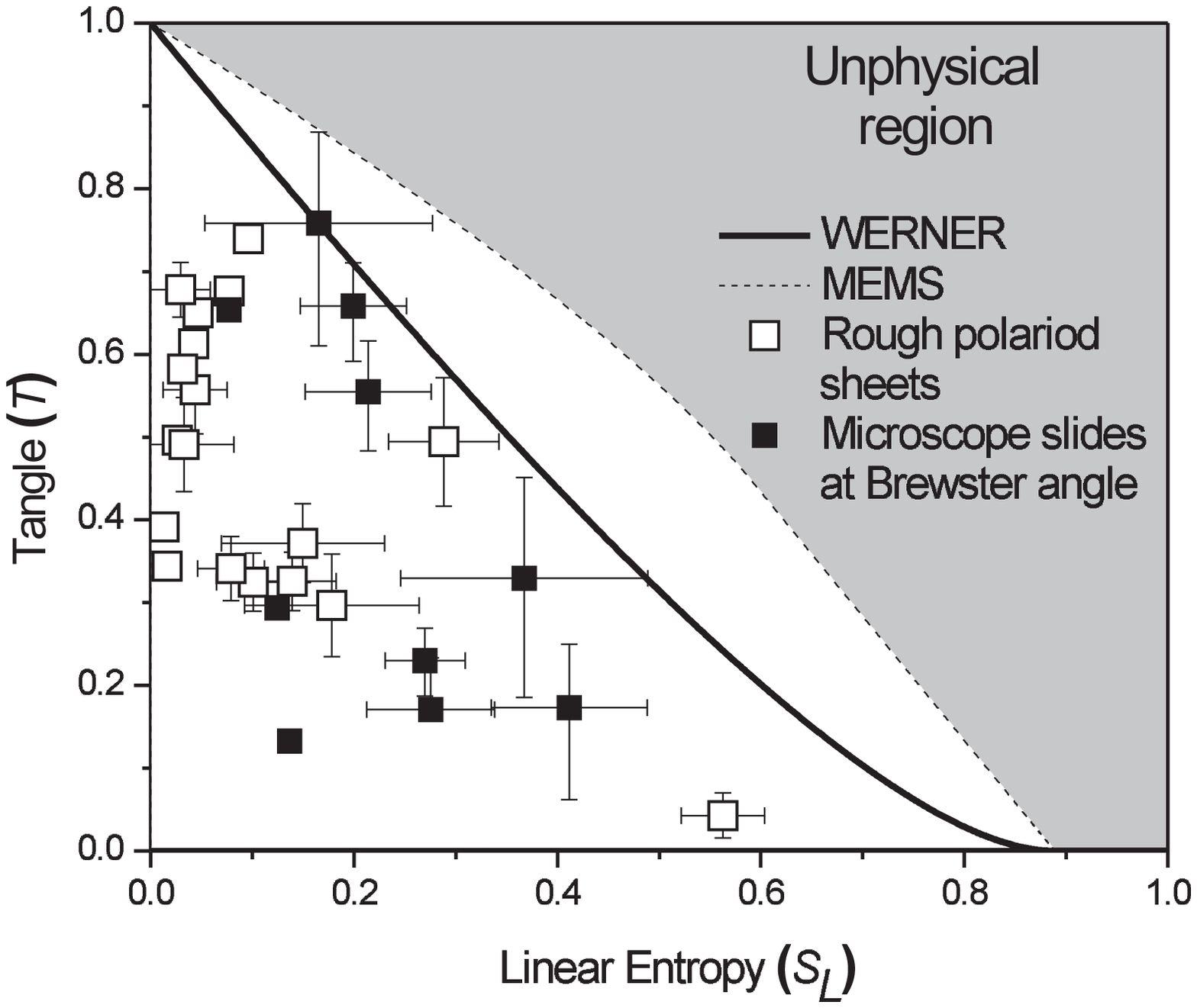}
\caption{\label{fig:3} Experimental data generated by dichroic
scattering systems (type III).}
\end{figure}

In Fig.~\ref{fig:3}, experimental data generated by dichroic
scattering systems (type III) are shown. We used:
 (i) Surface diffusers
followed by a stacks of microscope slides at the Brewster angle
and (ii) Commercially available polaroid sheets with
manually-added surface roughness on its front surface to provide
for wave-vector spread.
All data thus obtained fall below the Werner curve, generating
what we called sub-Werner states, namely states with a lower value
of  tangle ($T$) than a Werner state, for a given value of the
linear entropy ($S_L$).

In summary, Figs.~\ref{fig:2} (a)-(b) show that all data generated
by type I and II scattering systems fall on the Werner curve,
within the experimental error; while data generated by scattering
samples type III, which are presented in Fig.~\ref{fig:3}, lay
below the Werner curve. In Section III we shall present a simple
theoretical interpretation for such results.
\subsection{Error estimate}
In order to estimate the errors in our measured data, we
numerically generated 16 Monte Carlo sets $N_i$ ($i=1,\ldots,16$)
of $10^{3}$ simulated photon counts, corresponding to each of the
$16$ actual coincidence count measurements $\{n_{i}^\mathrm{exp}
\}$ ($i=1,\ldots,16$) required by tomographic analysis to
reconstruct a single two-photon density matrix.
Each set $N_i$ had a Gaussian distribution centered around the
mean value $\mu_{i} = n_{i}^\mathrm{exp}$, with standard deviation
$\sigma_{i}=\sqrt{n_{i}^\mathrm{exp}}$. The  sets $N_i$ where
created by using the ``NormalDistribution'' built-in function of
the program Mathematica 5.2.
Once we generated the $16$ Monte Carlo sets $N_i$, we
reconstructed the corresponding $10^3$ density matrices using a
maximum likelihood estimation protocol, to assure that they could
represent physical states. Finally, from this ensemble of matrices
we calculated the average tangle $T^\mathrm{av}$  and linear
entropy $S_{L}^\mathrm{av}$. The error bars were estimated as the
absolute distance between the mean quantities ($\mathrm{av}$) and
the measured  ones ($\mathrm{exp}$):
$\sigma_{T}=|T^\mathrm{exp}-T^\mathrm{av}|$,
$\sigma_{S_{L}}=|S_{L}^\mathrm{exp}-S_{L}^\mathrm{av}|$. It should
be noted that this procedure produces  an overestimation of the
experimental errors. In the cases where part of the overestimated
error bars fell into the unphysical region, the length of such
bars was limited to the border of the physically allowed density
matrices.
\subsection{Generalized Werner states}
Close inspection of the reconstructed density matrices generated
by type II scattering systems  revealed that in some cases the
measured states represented a generalized form of Werner states.
These are equivalent to the original Werner states $\rho_W$ with
respect to their values of $T$ and $S_L$, but the form of their
density matrices is different. Werner states $\rho_{W}$ of two
qubits were originally defined \cite{werner} as such states which
are $U \otimes U$ invariant: $\rho_{W}=U \otimes U \rho_{W}
U^{\dagger}\otimes U^{\dagger}$. Here $U \otimes U$ is any
symmetric separable unitary transformation acting on the two
qubits. The generalized Werner states $\rho_{GW}$ we
experimentally generated, can be obtained from $\rho_{W}$ by
applying a \emph{local} unitary operation $V$ acting upon only one
of the two qubits:
$\rho_{GW}=V\otimes I \rho_{W} V^{\dagger}\otimes I$, where $I=\left[\begin{array}{cc} 1 & 0\\
0 & 1 \\
    \end{array}\right]$, and
\begin{eqnarray}\label{first}
\label{eq:1}
V(\alpha,\beta,\gamma)=\left[\begin{array}{cc} e^{-\frac{i}{2}(\alpha+\beta)}\cos{\gamma/2} & -e^{-\frac{i}{2}(\alpha-\beta)}\sin{\gamma/2}\\
e^{\frac{i}{2}(\alpha-\beta)}\sin{\gamma/2} &
    e^{\frac{i}{2}(\alpha+\beta)}\cos{\gamma/2}\\
    \end{array}\right],
\end{eqnarray}
where $\alpha, \beta, \gamma$ can be identified with the three
Euler angles characterizing an ordinary rotation in $\mathbb{R}^3$
\cite{nielsen}. These generalized Werner states have the same
values of $T$ and $S_L$ as the original $\rho_{W}$ (since a local
unitary transformation does not affect neither the degree of
entanglement nor the degree of purity) but are no longer invariant
under unitary transformations of the form $U \otimes U$. By using
Eq. (\ref{first}), we calculated the average maximal fidelity of
the measured states $\rho^\mathrm{exp}_{GW}$ with the target
generalized Werner states
$\rho^\mathrm{th}_{GW}(p,\alpha,\beta,\gamma)$. We found
$\bar{F}(\rho^\mathrm{exp}_{GW},\rho^\mathrm{th}_{GW})\approx 96
\%$, revealing that our data are well fitted by this
four-parameter class of generalized Werner states.
\section{The phenomenological model}
\begin{figure}[h!]
\includegraphics[angle=0,width=9.5truecm]{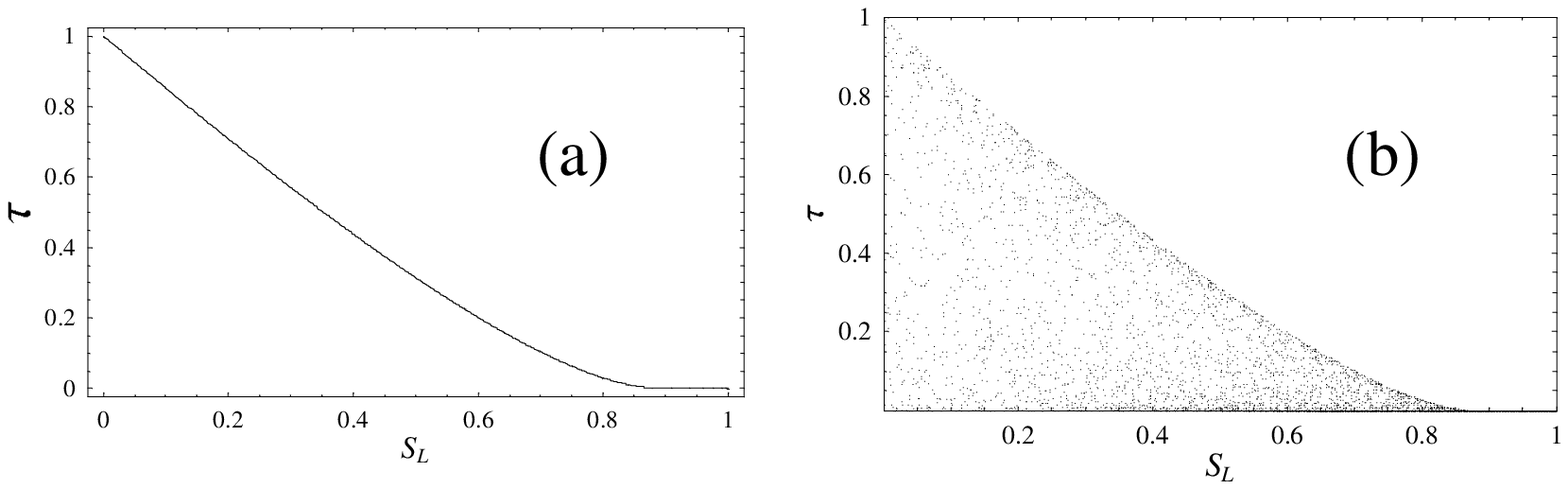}
\caption{\label{fig:4} Numerical simulation for our
phenomenological model. Fig. 4 (a) isotropic and birefringent
scattering, Fig. 4 (b) dichroic scattering.}
\end{figure}

In Ref. \cite{aiello3}, a theoretical study of the analogies
between classical linear optics and quantum maps was given. Within
this theoretical framework it is possible to build a simple
phenomenological model capable of explaining all our experimental
results.  To this end let us consider  the experimental set-up
represented in Fig. \ref{fig:1}. The linear optical scattering
element $\mathcal{T}_A$ inserted across path $A$ can be
\emph{classically} represented by some Mueller matrix
$\mathcal{M}$ \cite{hulst} which describes its
polarization-dependent interaction with a classical  beam of
light. However, $\mathcal{T}_A$ can also be represented by a
linear, completely positive, local \emph{quantum} map
$\mathcal{E}: \, \rho \rightarrow \mathcal{E}[\rho]$, which
describes  the interaction of the scattering element with a
two-photon light beam encoding a pair of polarization qubits.
These qubits are, in turn, represented by a $4 \times 4$ density
matrix $\rho$. Since $\mathcal{T}_A$ interacts with only one of
the two photons, the map $\mathcal{E}$ is said to be \emph{local}
and it can be written as $\mathcal{E} = \mathcal{E}_A \otimes
\mathcal{I}$, where $\mathcal{E}_A$ is the single-qubit (or
single-photon) quantum map representing $\mathcal{T}_A$, and
$\mathcal{I}$ is the single-qubit identity map.

It can be shown that the classical Mueller matrix $\mathcal{M}$
and the single-qubit quantum map $\mathcal{E}_A$ are univocally
related. Specifically, if with $\mathcal{M}$ we denote the
complex-valued Mueller matrix written in the standard basis, then
the following decomposition holds:
\begin{equation}\label{Mueller1}
\mathcal{M}=\sum_{\mu = 0}^3 \lambda_{\mu} T_{\mu} \otimes
 T_{\mu}^*,
\end{equation}
where $\{ T_\mu \}$ is a set of four $2 \times 2$ Jones matrices
\cite{hulst}, each representing a non-depolarizing linear optical
 element in classical polarization optics, and $\{ \lambda_\mu\}$ are the
four non-negative eigenvalues of the ``dynamical'' matrix $H$
associated to $\mathcal{M}$. Given Eq. (\ref{Mueller1}), it is
possible to show that the two-qubit quantum map $\mathcal{E}$ can
be written as
\begin{eqnarray}\label{deco}
\rho_\mathcal{E} = \mathcal{E}[\rho]  \propto \sum_{\mu = 0}^3
\lambda_\mu \,
 T_\mu \otimes I \, \rho \,    T_\mu^\dagger
\otimes I,
\end{eqnarray}
where the proportionality symbol ``$\propto$'' on the right hand
side of Eq.~(\ref{deco}) accounts for a possible renormalization
to ensure $\tr (\rho_\mathcal{E})=1$. Such renormalization becomes
necessary when $\mathcal{T}_A$ presents polarization-dependent
losses (i.e., dichroism). We anticipate that when such
re-normalization is necessary the map is considered non-trace
preserving. We shall briefly discuss this issue in the conclusion.

With these ingredients, a phenomenological polarization-scattering
model can be built as follows. First we use the polar
decomposition \cite{Lu96} to write an arbitrary Mueller matrix
$\mathcal{M}=\mathcal{M}_{\Delta}\mathcal{M}_{B}\mathcal{M}_{D}$,
where $\mathcal{M}_{\Delta}$, $\mathcal{M}_{B}$ and
$\mathcal{M}_{D}$ represent a purely depolarizing element, a
birefringent (or retarder) element, and a dichroic (or
diattenuator) element, respectively.  Specific analytical
expressions for $\mathcal{M}_{\Delta}$, $\mathcal{M}_{B}$ and
$\mathcal{M}_{D}$ can be found in the literature \cite{kliger}.
Second, we use Eq.~(\ref{Mueller1}) to find the quantum maps
corresponding to $\mathcal{M}_{\Delta}$, $\mathcal{M}_{B}$ and
$\mathcal{M}_{D}$ and, by using such maps, we calculate the
scattered two-photon state $\rho_{\mathcal{E}}$. In our
experimental realizations we used isotropic scatterers
$\mathcal{M}_{IS}=\mathcal{M}_{\Delta}$ with isotropic
depolarization factor $0 \leq \Delta < 1$, birefringent scattering
media $\mathcal{M}_{BS}$, described in terms of the product of a
purely birefringent medium $\mathcal{M}_{B}$ and an isotropic
depolarizer $\mathcal{M}_{\Delta}$, i.e.
$\mathcal{M}_{BS}=\mathcal{M}_{B}\mathcal{M}_{\Delta}$, and
finally, dichroic scattering media
$\mathcal{M}_{DS}=\mathcal{M}_{D}\mathcal{M}_{\Delta}$, which are
in turn described by a product of a purely dichroic medium
$\mathcal{M}_{D}$ and a purely depolarizing medium
$\mathcal{M}_{\Delta}$. It should be noted that these product
decompositions are not unique. Other decompositions with different
orders are possible but the elements of each matrix might change,
since the matrices $\mathcal{M}_{\Delta}$, $\mathcal{M}_{B}$ and
$\mathcal{M}_{D}$ do not commute.

Filling in the above expressions with random numbers selected from
suitably chosen ranges, we simulated all scattering processes
occurring in our experiments. Fig.~\ref{fig:4} shows a numerical
simulation of the scattered states in the tangle vs. linear
entropy plane, obtained with the singlet two-photon state as input
state. Fig. \ref{fig:4}~(a) corresponds to isotropic and
birefringent scatterers,  and Fig.~\ref{fig:4}~(b) to dichroic
scatterers. The qualitative agreement between this model and the
experimental results shown in Fig.~\ref{fig:2} and
Fig.~\ref{fig:3} is manifest.

\section{Conclusions}
In summary, we have presented experimental results on entanglement
properties of scattered photon-pairs for three varieties of
optical scattering systems. In this way we were able to generate
two distinct types of two-photon mixed states; namely Werner-like
and sub-Werner-like states. Moreover, we have introduced a simple
phenomenological model based onto the analogy between classical
polarization optics and quantum mechanics of qubits, that fully
reproduces our experimental findings. In the case of sub-Werner
states, the phenomenological model represents a non-trace
preserving quantum map. Such description might be considered
controversial since a non-trace preserving local map  can in
principle lead to violation of causality when it describes the
evolution of a composite system made of two spatially separate
subsystems \cite{Aiello062}. However, we argue that our measured
states do not violate the signaling condition as they are
post-selected by the coincidence measurement, a procedure which
involves classical communication between the two detectors.
Finally, we expect it to be possible to create states \emph{above}
the Werner curve (in particular MEMS) \cite{mems2,mems3}, by
post-selective detection when acting on a single photon
\cite{Aiello062}. Work along this line
is in progress in our group.\\

\begin{acknowledgments}
This project is part of the program of FOM and is also supported
by the EU under the IST-ATESIT contract.

We gratefully acknowledge M. B. van der Mark for making available
the bundle of parallel fibers \cite{vdmark}.
\end{acknowledgments}

\end{document}